\documentclass[sigconf]{acmart}

\renewcommand\footnotetextcopyrightpermission[1]{} % removes footnote with conference information in first column
\usepackage{booktabs} % For formal tables
\usepackage{graphicx}
\usepackage{subfigure}

\usepackage{algorithm}
\usepackage{algorithmic}

\setcopyright{rightsretained}
\author{Chuanchang Chen, Yubo Tao, Hai Lin}

\newtheorem{theorem}{Definition}[section]

\begin{document}
\title{Dynamic Network Embeddings for Network Evolution Analysis}
%\titlenote{Produces the permission block, and
%  copyright information}
%\subtitle{Extended Abstract}
%\subtitlenote{The full version of the author's guide is available as
%  \texttt{acmart.pdf} document}

\begin{abstract}
Network embeddings learn to represent nodes as low-dimensional vectors to preserve the proximity between nodes and communities of the network for network analysis. The temporal edges (e.g., relationships, contacts, and emails) in dynamic networks are important for network evolution analysis, but few existing methods in network embeddings can capture the dynamic information from temporal edges. In this paper, we propose a novel dynamic network embedding method to analyze evolution patterns of dynamic networks effectively. Our method uses random walk to keep the proximity between nodes and applies dynamic Bernoulli embeddings to train discrete-time network embeddings in the same vector space without alignments to preserve the temporal continuity of stable nodes. We compare our method with several state-of-the-art methods by link prediction and evolving node detection, and the experiments demonstrate that our method generally has better performance in these tasks. Our method is further verified by two real-world dynamic networks via detecting evolving nodes and visualizing their temporal trajectories in the embedded space.
\end{abstract}




\maketitle

\section{INTRODUCTION}

Networks describe the relationships between entities, such as social relationships within a population, interactions between biological proteins, and co-occurrence relationships within words. Dynamic networks are very common in many application domains with entities and connections changing over time, such as email networks and instant messaging networks. Analyzing these networks can gain insight into evolution patterns in these domains. Recently, dynamic network analysis has received much attention, and many embedding-based approaches have been proposed for temporal link prediction~\cite{CTDNE}, node prediction~\cite{triad}, and multi-label node classification~\cite{DepthLGP}. Thus, utilizing embeddings to investigate network evolution is a promising research direction.

\begin{figure}
\centering
\includegraphics [width=8.4cm]{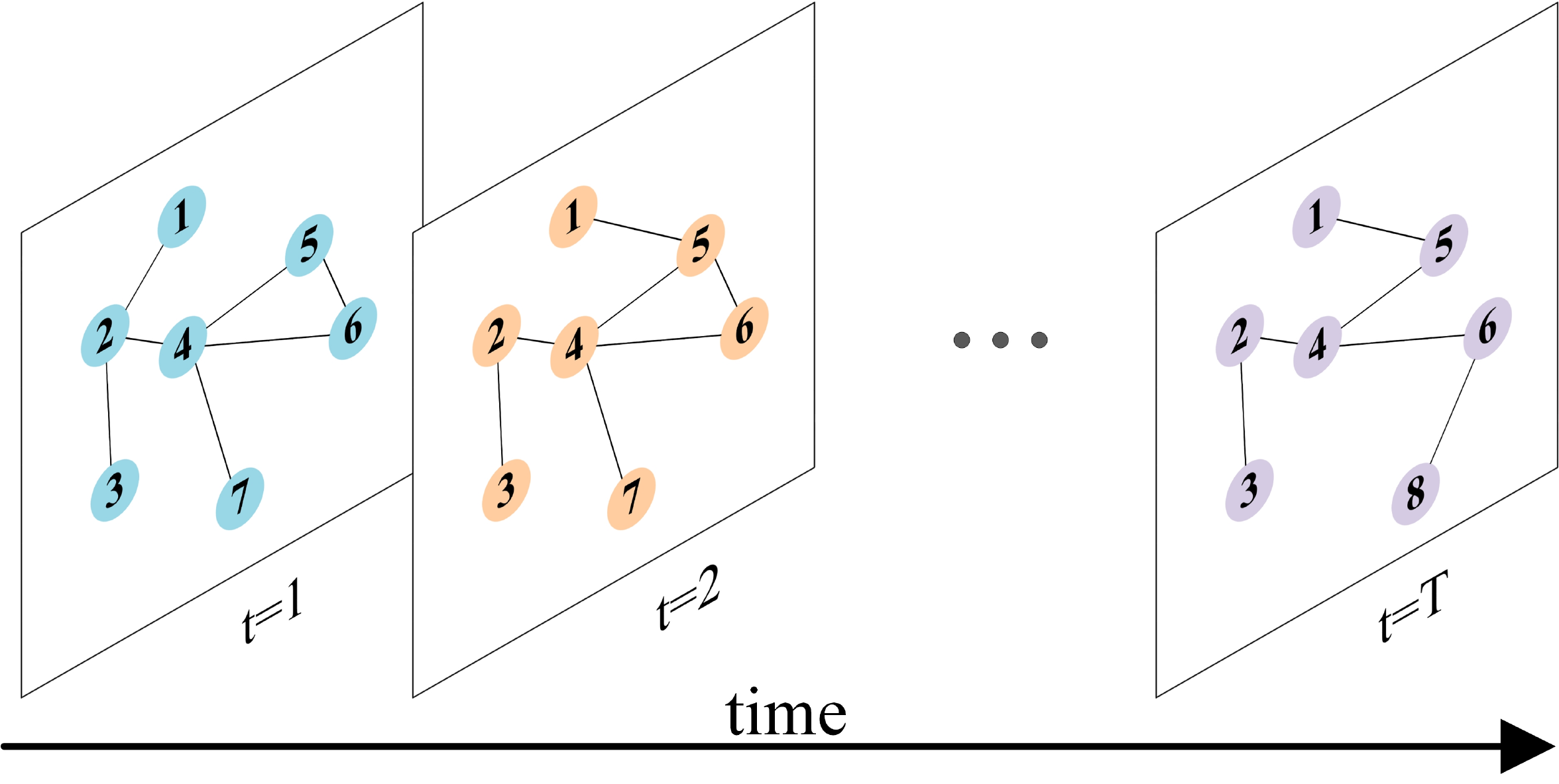}
\caption{An illustration of a social network with the addition and deletion of nodes and edges over time. For example, node 7 disappears at $t = T$, and node 8 appears at $t = T$. The edge between node 1 and node 2 is deleted at $t = 2$, and a new connection between node 1 and node 5 is generated.}
\label{fig:example}
\end{figure}

As illustrated in Figure~\ref{fig:example}, a social network is evolving with the addition and deletion of nodes and edges over time. Each node represents a person, and each edge indicates one kind of events between two persons, such as friendship, contacts, and emails. The weight of an edge denotes how strong this relationship is or the frequency of communication (e.g., the number of emails or calls). Two most important factors in dynamic networks are the proximity between nodes and temporal continuity of stable nodes along the time, which should be preserved in network embeddings for effective network evolution analysis. For instance, node 1 and node 3 share the same neighbor nodes at $t = 1$ in Figure~\ref{fig:example}, but this relation is decreased at $t = 2$ (i.e., proximity similarity). The neighbor nodes of node 4 at $t = 1$ and $t = 2$ are similar. Thus, the embedded vectors of node 4 at the two timesteps should be nearly the same (i.e., temporal similarity).

Previous methods for network embeddings, such as matrix factorization~\cite{GraRep}, random walk~\cite{deepwalk,node2vec,struc2vec}, deep neural network~\cite{DNNF,SDNE,CRDM,ANE}, and many others~\cite{LINE,RaRE}, generally focus on static networks to preserve the proximity between nodes. These methods fail to capture the temporal information in dynamic networks. Recently, continuous-time dynamic network embeddings~\cite{CTDNE} aim to learn continuous changes in temporal networks by temporal walk, which is a walking strategy with a time-ascending order. However, this method represents all network information into one embedding and cannot effectively capture the temporal changes of nodes over time, such as evolving node detection. DynamicTriad~\cite{triad} seeks to train all graphs in a dynamic network jointly into a sequence of embeddings by imposing \textit{triad}. However, the triad, a local proximity, only describes the local relation among up to three nodes, and can hardly capture high-order proximities between nodes.

%TNE~\cite{TNE} assumes that the changes in the dynamic network appear at a short duration to predict the future connections via the previous networks.
%DynGEM~\cite{DynGEM} leverage a deep learning model to learn dynamic network incrementally.

In this paper, we attempt to learn discrete-time network embeddings for dynamic network, i.e., each node has the same number of low-dimensional vectors with the number of timesteps and both the proximity and temporal continuity of each node are preserved in the dynamic network embeddings. A direct method to solve this problem is to learn each graph separately by static network embedding methods, and then align all network embeddings into the same vector space by alignment methods~\cite{DWER}. However, it is challenging to align these embeddings due to nonlinear movements of evolving nodes, and such alignment error would reduce the performance of downstream tasks. Thus, we propose a novel dynamic network embedding method to learn a sequence of graphs in a dynamic network and generate continuous embedded vectors for stable nodes without embedding alignments. Dynamic network embeddings can capture neighbor node changes over time for evolution analysis. Generally, this paper has the following contributions:
\begin{itemize}
\item We propose a general dynamic network embedding method that incorporates random walk on dynamic networks into Bernoulli embeddings.
\item Our method is more effective in link prediction when compared with other state-of-the-art techniques. Besides, we generate artificial dynamic networks to verify our method in capturing the temporal evolution of nodes and achieve the best performance.
\item Our method can be used to analyze and visualize the trajectories of evolving nodes while preserving the temporal continuity of stable nodes over time.
\end{itemize}

\section{RELATED WORK}

With the rise of social networks (e.g., Facebook and Twitter) and big data (e.g., millions or billions of interaction records), network embedding methods have received considerable attention from both industrial and academic. The key point of network embeddings is to learn a low-dimensional vector representation for each node to preserve the proximity, which can be easily used for several application tasks, for instance, node classification~\cite{deepwalk}, link prediction~\cite{ATPG}, node clustering~\cite{CPNE}, anomaly detection~\cite{AEAT}, and collaboration prediction~\cite{TGAP}.

Some network embedding methods are based on matrix factorization~\cite{GraRep}, which constructs a $k$-step transition probability matrix to measure the node similarity at different scales. Inspired by a good performance of word2vec in natural language processing, some researchers incorporated random walk into the skip-gram model~\cite{EEOW} to learn network embeddings, such as DeepWalk~\cite{deepwalk} and node2vec~\cite{node2vec}. These methods use random walk to produce a series of node sequences and apply the skip-gram model to learn network representations. Struc2vec~\cite{struc2vec} focuses on the structural identities of nodes and constructs a weighted multilayer graph for random walk to capture the hierarchical structural similarity. Recently, some embedding methods based on deep neural networks have received considerable attention~\cite{SDNE,DNNF,CRDM} to learn nonlinear mapping functions. To enhance the robustness of representations, Dai et al. employed generative adversarial network to capture latent features in network embeddings~\cite{ANE}.

Most previous network embedding methods only focus on static networks, but dynamic network embedding learning is a hot research topic. It is related to dynamic latent space models, such as a dynamic model accounting for friendships drifting over time~\cite{DSNA} and a case-control approximate likelihood~\cite{FIFT}. CTDNE~\cite{CTDNE} incorporates temporal information into existing network embedding methods based on random walk by introducing a time-series order. TNE~\cite{TNE} is a discrete-time dynamic network embedding method based on matrix factorization. DynGEM~\cite{DynGEM} is based on deep autoencoders combined with a layer expansion to generate embeddings of a growing network. DynamicTriad~\cite{triad} focuses on the local structure called triad to learn the proximity information and evolution patterns. TIMERS~\cite{TIMERS} uses a SVD model to learn dynamic network embeddings incrementally based on the initialization of the previous graph. DepthLGP~\cite{DepthLGP} tackles the issue of updating out-of-sample nodes into network embeddings by combining a probabilistic model with deep learning. Previous methods evaluate network embeddings only by static tasks, and the trajectories of evolving nodes have not received much attention. Therefore, one of our evaluations is evolving node detection, and we further visualize the trajectories of evolving nodes in the context of stable nodes for evolution pattern analysis.

\section{PROBLEM DEFINITION}

In this paper, we seek to solve the proximity and temporal representation problem of dynamic networks, i.e., each node has one embedded vector in each timestep. Recently, the exploration of word meaning evolution in natural language processing has received much attention, and the key is to understand how words change their meanings over time and mine the latent cultural evolution~\cite{semanticsurvey}. Kulkarni et al. proposed an insightful conception of aligning all word embeddings at different timesteps into one vector space before semantic shift analysis~\cite{kulkarni}. Instead of a linear transformation for the alignment, Eger and Mehler presented second-order embeddings to compare the difference of word meanings~\cite{eger}. Moreover, it was shown in \cite{bamler} and \cite{yao} that we can learn diachronic word embeddings in the same vector space jointly. Thus the alignment across of embeddings is simultaneous and accurate.

Inspired by these diachronic word embeddings, we propose a novel method to generate embeddings for dynamic networks. The definition of dynamic networks is given as follows:
\begin{theorem}[Dynamic Networks]
A dynamic network is a series of graphs $\Gamma=\{G_1,...,G_T\}$ and $G_t=(V_t,E_t)$, where $T$ is the number of graphs, $V_t$ is a node set and $E_t$ includes all temporal edges within the timespan $[S_t,S_{t+1}]$. Each $e_i=(u,v,s_i)\in E_t$ is a temporal edge between the node $u\in V_t$ and the node $v \in V_t$ at the timestamp $s_i\in[S_t,S_{t+1}]$.
\end{theorem}

A dynamic network can be constructed from temporal events, and different construction methods may have different applications, as discussed in Section~\ref{sec:construction}. Our goal is to learn dynamic network embeddings $M_1,...,M_T$ and $M_t\in \mathbb{R}^{|V_t|\times D}$, where $|V_t|$ is the number of nodes at timestep $t$ and $D$ is the dimension of embeddings. Thus, the concept of dynamic network embeddings is defined formally as follows:
\begin{theorem}[Dynamic Network Embeddings]
Given a dynamic network $\Gamma=\{G_1,...,G_T\}$, dynamic network embeddings aim to project a node $g\in V_t$ into a low-dimensional vector space by a mapping function $\textit{f}:g\mapsto y_g^{(t)}\in \mathbb{R}^D, D\ll\max|V_t|,t\in[1,T]$.
\end{theorem}

Thus, there is an embedding matrix $M_t\in \mathbb{R}^{|V_t|\times D}$ to represent the proximity and temporal properties for each graph $G_t$. The dynamic network embeddings generally require the following characteristics:
\begin{itemize}
\item \textbf{Proximity Preservation}. The embeddings should preserve the proximity between nodes, i.e., if a node $u$ and a node $v$ have similar neighbor nodes at timestep $t$, $y_u^{(t)}$ and $y_v^{(t)}$ should be located nearby in the vector space.
\item \textbf{Temporal Continuity}. The embeddings should keep the temporal similarity of stable nodes, i.e., if a node $u$ has similar neighbors at timestep $t$ and $t+1$, $y_u^{(t)}$ and $y_u^{(t+1)}$ should be located nearby in the vector space.
\item \textbf{Dimension Reduction}. Although dynamic networks can be complex with thousands of nodes and temporal edges, the embeddings should be low-dimensional, i.e., $D \ll max|V_t|$. Therefore, the embeddings can be effectively applied to downstream machine learning tasks.
\end{itemize}

\section{PROPOSED METHOD}

Our method has three main steps. Firstly, we construct a dynamic network from temporal events. Then we apply random walk to generate node sequences matrix for each graph to keep the local proximity for each node in each timestep. Finally, we learn node representations from all node sequences of all timesteps jointly based on Bernoulli embeddings, as shown in Algorithm~\ref{alg1}.

\subsection{Dynamic Network Construction}
\label{sec:construction}

The relationships between entities are generally described by timestamped events in real-world datasets, such as emails, calls, and interaction records. We first need to transform these temporal events into a dynamic network by constructing a sequence of graphs before learning its dynamic network embeddings.

One strategy is the fixed time interval $\omega$ (e.g., hours, days, and weeks) for each timestep. Thus, each graph $G_t$ has a time window $[S_t, S_t + \omega)$ and $S_t$ is the earliest timestamp of timestep $t$. The temporal edge set of timestep $t$ is
\begin{equation}
    E_t=\{e_i|s_i\in[S_t,S_t+\omega)\}.
\end{equation}
Since the events may be not evenly distributed over time, graphs would have different numbers of events. For example, one graph may contain one thousand temporal edges, while the other graph may only have one hundred temporal edges. The dynamic network embeddings learned from these graphs with non-uniform events may have a negative impact on downstream tasks, especially for sparse graphs. Thus, it would be better to choose different time intervals for different timesteps, and construct a dynamic network by a fixed number of events $\varepsilon$ (the other strategy). The events can be generally described as follows (the number of events is $N$):
\begin{equation}
    A=(e_1,e_2,...e_N).
\end{equation}
Each graph $G_t$ has an event window $(\varepsilon\cdot(t-1),\varepsilon\cdot t]\subset [1,N]$ , and the edge set of timestep $t$ is defined as follows:
\begin{equation}
    E_t=\{e_i|i\in(\varepsilon\cdot(t-1),\varepsilon\cdot t]\}.
\end{equation}
In this way, each graph has nearly the same number of events. However, the drawback is breaking the equivalent time interval, which may be important for some applications. We can select the fixed time interval or the fixed number of events depending on tasks when constructing dynamic networks.

\begin{figure}
\centering
\includegraphics [width=8cm]{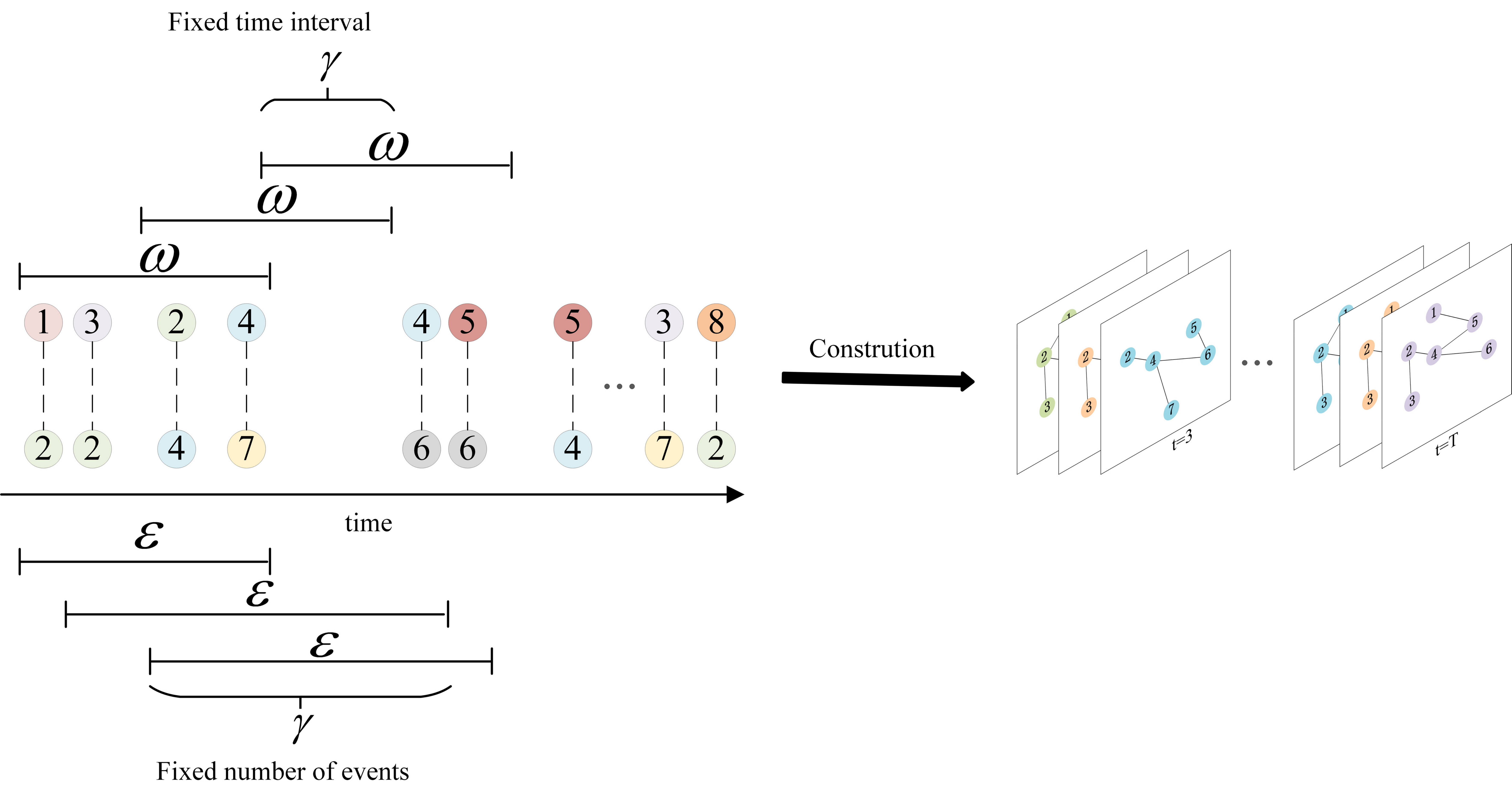}
\caption{Dynamic network construction by the fixed time interval or the fixed number of events.}
\label{fig:construction}
\end{figure}

The hard boundary may generate discontinuous network embeddings over time and lead to misleading evolution patterns. To address this issue, the window of each graph has an overlap $\gamma\in[0,1]$ with the previous graph, as shown in Figure~\ref{fig:construction}. For the fixed time interval, the time window $\omega>\Delta t$ (non-overlapping time interval) and the overlap $\gamma$ is defined as $\gamma=(\omega-\Delta t)/\omega$.
For the fixed number of events, the event window $\varepsilon>\Delta e$ (non-overlapping events) and the overlap $\gamma$ is defined as $\gamma=(\varepsilon-\Delta e)/\varepsilon$.
In our experiment, we choose to adjust $\varepsilon$ for the link prediction and evolving node detection tasks, and adjust $\omega$ for analyzing and visualizing the trajectories of evolving nodes.

\begin{algorithm}[htb]
\caption{Dynamic Network Embeddings}
\label{alg1}
\begin{algorithmic}
\REQUIRE
A dynamic networks $\Gamma=\{G_1,...,G_T\}, G_t=(V_t, E_t)$\\
number of walks on each node $r$\\
length of each walk $L$\\
embedding size $D$\\
context size $cs$\\
negative sample size $ns$\\
\ENSURE Network embeddings $M_1,...,M_T$\\
Initialize embedding matrix $M_1,...,M_T$\\
Initialize context matrix $\alpha$\\
%$V_t$ is the set of nodes appearing at $G_t$\\
%$y_g^{(t)}\in M_t$, $\alpha_g\in\alpha$
\FOR{$t=1$ to $T$}
\FOR{each $g\in V_t$}
\STATE $W_g^{(t)}=RandomWalk(G_t,g,r,L)$
%=\{(v_1,...,v_L),...,(v_{L\cdot(r-1)},...,v_{L\cdot r})\}$
\STATE DBE($y_g^{(t)}$,$\alpha_g$,$W_g^{(t)}$,$V_t$,$cs$,$ns$)
\ENDFOR
\ENDFOR
\end{algorithmic}
\end{algorithm}

\subsection{Sequential Random Walks}

We use random walk to capture the proximity information of each graph and generate sequences of walks as the input for network embeddings learning. For each graph, we first choose a node as the root of a random walk, then the walk selects uniformly from the neighbors of the last node visited until the maximum length $L$ of a node sequence is reached. The walking process will repeat $r$ times for each node of each graph. Instead of repeating the walking process in each graph iteratively, our approach runs random walk on graphs parallelly to generate one node sequence matrix $W_t$ for each graph $G_t$. In natural language processing, the \textit{context} is composed of the words appearing to both the right and left of the given word. For the network, the \textit{context} means the neighbors of the node, and the nodes before and after the node in the node sequence are the context of the node.

\subsection{Dynamic Bernoulli Embeddings}

We introduce the dynamic Bernoulli embedding method for latent representation learning of a dynamic networks (Algorithm~\ref{alg2}). A node sequence $w_j^{(t)}=\{v_1^{(t)},...,v_L^{(t)}\}$ is generated from the node $j$ at timestep $t$ by random walk. We define the context of $v_i^{(t)}$ in the node sequence as $c_i^{(t)}=\{v_{i-cs/2}^{(t)},...,v_{i+cs/2}^{(t)}\}\setminus v_i^{(t)}$, where the context size is $cs$. $x_{i}^{(t)}\in \mathbb{R}^{|V_t|}$ is an indicator vector that each entry $x_{ig}$ is zero or one, and $x_{ig}^{(t)}=1$ means $v_i^{(t)}$ is the node $g$.

We assign $x_{ig}^{(t)}$ with a conditional distribution $P(x_{ig}^{(t)} | c_i^{(t)})$ based on a Bernoulli probability. Likewise, we define $\alpha_g\in \mathbb{R}^D$ as the context for the node $g$ and we update it in the process of dynamic embedding learning at each timestep. In contrast, we only update the embedding vector $y_g^{(t)}$ of $M_t$ when we train the node $v_i^{(t)}$ at timestep $t$. Finally, we define the representation function of $g$ and $c_i^{(t)}$ at timestep $t$ as follows:
\begin{equation}
    \eta_{ig} = y_{g}^{(t)T}(\sum_{\substack{k\neq i \\ k\in[i-cs/2,i+cs/2]}}\sum_{g\in V_t}\alpha_{g} x_{k{g}}^{(t)}),
\label{equ:etaig}
\end{equation}
where the sum $\sum_{g}\alpha_{g} x_{kg}^{(t)}$ is a filter to select the context vector $\alpha_g$. With the shared context vector $\alpha_g$ for all timesteps, we can train all embeddings in the same vector space and capture the evolution information across all timesteps.

To initialize embeddings $M_1,...,M_T$ and $\alpha$, we use Gaussian priors with a diagonal covariance, and set parameters $\lambda_1$ and $\lambda$ according to \cite{DBE}.
\begin{align}
    \alpha_g,y_g^{(1)}&\sim N(0,\lambda_1^{-1}I)\\
    y_g^{(t)}&\sim N(y_g^{(t-1)},\lambda^{-1}I).
\end{align}
To achieve better performance, we can use embeddings pretrained by static methods as the default value of $M_t$ to speed up the convergence of our model.

To train dynamic network embeddings, we regularize the \textit{pseudo log likelihood} with the log priors, and then maximize the likelihood to obtain a pseudo MAP estimation. Furthermore, we consider the contributions of positive context nodes and negative context nodes separately, and we define these two likelihoods $\mathcal{L}_{pos}$ and $\mathcal{L}_{neg}$ as follow:
\begin{align}
    \label{equ:Lpos}\mathcal{L}_{pos} & =\sum_{i=1}^{L}\sum_{g\in{V_t}}x_{ig}^{(t)}\log \sigma(\eta_{ig}),\\
    \label{equ:Lneg1} \ \mathcal{L}_{neg} & =\sum_{i=1}^{L}\sum_{g\in{V_t}}(1-x_{ig}^{(t)})\log(1-\sigma(\eta_{ig})),
\end{align}
where $\sigma(\cdot)$ is the sigmoid function to generate the probability. Considering negative context nodes are far more than positive context nodes, we use negative sampling to randomly select nodes except node $g$ as negative context nodes to reduce the computation of $\mathcal{L}_{neg}$. We define the sampling distribution of negative context nodes is $\phi$, and Equation~\ref{equ:Lneg1} is redefined as:
\begin{equation}
    \label{equ:Lneg2} \ \mathcal{L}_{neg} =\sum_{i=1}^{L}\sum_{g\sim\phi}\log(1-\sigma(\eta_{ig})).
\end{equation}
In this paper, we set the unigram distribution~\cite{unigram} raised to the power of 0.75 as $\phi$. We notice that $y$ and $\alpha$ are the terms of Equation~\ref{equ:etaig} which is included in Equation~\ref{equ:Lpos} and~\ref{equ:Lneg2}. Thus, we set a prior to $\alpha$:
\begin{equation}
    \mathcal{L}_{\alpha}=-\frac{\lambda_1}{2}\sum_{g\in{V_t}}\|\alpha_g\|^2.
\end{equation}
To penalize the consecutive embedding vector $y_g^{(t-1)}$ and $y_g^{(t)}$ for drifting from each too far apart, we set the prior of $y$ as:
\begin{equation}
    \label{equ:Ly}\mathcal{L}_{y}=-\frac{\lambda_1}{2}\sum_{g\in {V_t}}\|y_g^{(1)}\|^2-\frac{\lambda}{2}\sum_{\substack{g\in{V_t} \\ t\in[1,T]}}\|y_g^{(t)}-y_g^{(t-1)}\|^2.
\end{equation}
Finally, we group all likelihoods as the optimization objective:
\begin{equation}
    \label{equ:Lya}\mathcal{L}(y,\alpha)=\mathcal{L}_{pos}+\mathcal{L}_{neg}+\mathcal{L}_{\alpha}+\mathcal{L}_{y}.
\end{equation}
Overall, we use Stochastic Gradient Descent (SGD)~\cite{ASAM} to fit Equation~\ref{equ:Lya} with a proper learning rate.
\begin{algorithm}[htb]
\caption{DBE($y_g^{(t)}$,$\alpha_g$,$W_g^{(t)}$,$V_t$,$cs$,$ns$)}
\label{alg2}
\begin{algorithmic}
\FOR{each $w_k^{(t)}\in W_g^{(t)}$}
\FOR{each $v_i\in w_k^{(t)}$}
\STATE $V_{v_i}=NegativeSampling(V_t,v_i,ns)$
\STATE Minimize loss $\mathcal{L}(y,\alpha)$ by SGD($y_g^{(t)}$,$\alpha_g$,$V_t$,$cs$)
\STATE Update $y_g^{(t)}$ and $\alpha_g$
\ENDFOR
\ENDFOR
\end{algorithmic}
\end{algorithm}

\section{EXPERIMENTS}

Our method is evaluated by the link prediction task and the evolving node detection task. The former is a classical method to assess the effectiveness in capturing dynamic changes of the proximity in adjacent timesteps. The latter focuses on detecting evolving nodes which are unstable and likely to change over time.

For the first task, we use eight datasets collected from Network Repository~\cite{TNDR} and all datasets are temporal and real. Table~\ref{tab:datasets} shows the statistics of these datasets.

For the second task, we generate several artificial dynamic networks. Each dynamic network has different numbers of nodes and edges. For the density of networks, we impose a \textit{power-law distribution} on the node degree with different parameters.
\begin{equation}
    degree\propto CX^{-\alpha}.
\label{equ:power}
\end{equation}
There are 4 communities initially and the edges within the community are more than the edges between communities. Furthermore, to simulate the evolving trend of dynamic networks, we randomly choose 10\% nodes in the network as evolving nodes and design the evolution strategies of nodes as follows:
\begin{itemize}
\item Evolving nodes can change more than two edges at each timestep, while the limit is two for stable nodes.
\item Evolving nodes are gradually moved to another community by decreasing the number of edges within the community and increasing the number of edges with another community. The edges of stable nodes can be changed within the community.
\item The number of edges is generally stable for evolving nodes, i.e., the number of additions is nearly the same with the number of deletions.
\end{itemize}

\begin{table}
\footnotesize
  \caption{The statistics of dynamic networks. $|V|$ = number of nodes; $|E_T|$ = number of temporal edges; \={d} = average node degree in all timesteps; $d_{max}$ = max node degree in all timesteps; $S_t$ = the whole timespan (days); $T$ = number of timesteps in the training data.}
  \label{tab:datasets}
  \begin{tabular}{rrrrrrr}
    \toprule
    Dataset&$|V|$&$|E_T|$&\={d}&$d_{max}$&$S_t$&$T$\\
    \midrule
    IA-CONTACT &274 &28.2K &16 &113 &3.97 &10\\
    IA-HYPER &113 &20.8K &7.6 &77 &2.46 &14\\
    IA-ENRON &151 &50.5K &4.2 &61 &1137.55 &36\\
    IA-RADOSLAW-EMAIL &167 &82.9K &20.8 &239 &271.19 &19\\
    IA-EMAIL-EU &987 &332.3K &16.1 &232 &803.93 &44\\
    FB-FORUM &899 &33.7K &10.8 &109 &164.49 &10\\
    SOC-SIGN-BITCOINA &3783 &24.1K &6.4 &597 &1901.00 &11\\
    SOC-WIKI-ELEC &6271 &107K &13.1 &602 &1378.34 &20\\
  \bottomrule
\end{tabular}
\end{table}

\begin{table*}
  \caption{AUC scores of the link prediction task.}
  \label{tab:link}
  \begin{tabular}{rccccccc}
    \toprule
    Dataset&\textbf{DeepWalk}&\textbf{Node2vec}&\textbf{CTDNE}&\textbf{TNE}&\textbf{DynGEM}&\textbf{DynamicTriad}&\textbf{Our method}\\
    \midrule
    IA-CONTACT &0.845 &0.874 &0.913&0.880 &0.907&0.939 &\textbf{0.951}\\
    IA-HYPER &0.620 &0.641 &0.671&0.710 &0.736&0.792 &\textbf{0.816}\\
    IA-ENRON &0.719 &0.759 &0.877&0.822 &0.845&0.902 &\textbf{0.935}\\
    IA-RADOSLAW-EMAIL &0.734 &0.741 &0.811&0.831 &0.788&0.764 &\textbf{0.845}\\
    IA-EMAIL-EU &0.820 &0.860 &0.890&850 &0.864&\textbf{0.907} &0.878\\
    FB-FORUM &0.670 &0.790 &0.826&0.810 &0.856&0.825 &\textbf{0.920}\\
    SOC-SIGN-BITCOINA &0.840 &0.870 &0.891&0.877 &0.879&0.881 &\textbf{0.895}\\
    SOC-WIKI-ELEC &0.820 &0.840 &0.857&0.837 &0.822&0.849 &\textbf{0.859}\\
  \bottomrule
\end{tabular}
\end{table*}

\begin{table*}
  \caption{MAP, MRR and TOP-K scores of the evolving node detection task. $|V_{hot}|$ = number of evolving nodes at each timestep.}
  \label{tab:detection}
  \begin{tabular}{r|c|cccccccc}
    \toprule
    Parameters setting&Metric&\textbf{DeepWalk}&\textbf{Node2vec}&\textbf{TNE}&\textbf{DynGEM}&\textbf{DynamicTriad}&\textbf{Our method}&$|V_{hot}|$&$|E_T|$\\
    \midrule
    $\alpha=2$ &MAP &0.451 &0.453 &0.764&0.820&0.869&\textbf{0.900}&&\\
    $C=10^2$ &MRR &0.193 &0.203 &0.225&0.269&0.259&\textbf{0.281} &50 &5,010\\
    $N$=500 &TOP-K &0.389 &0.389 &0.733&0.766&0.789&\textbf{0.822} &&\\
    \hline
    $\alpha=2$ &MAP &0.551 &0.552 &0.689&0.754&0.779&\textbf{0.844}&&\\
    $C=10^3$ &MRR &0.225 &0.224 &0.240&0.260&0.254&\textbf{0.273} &50 &10,650\\
    $N$=500 &TOP-K &0.511 &0.489 &0.666&0.664&0.711&\textbf{0.789} &&\\
    \hline
    $\alpha=3$ &MAP &0.515 &0.498 &0.656&0.756&0.733&\textbf{0.795}&&\\
    $C=10^2$ &MRR &0.220 &0.215 &0.230&0.251&0.244&\textbf{0.267} &50 &20,838\\
    $N$=500 &TOP-K &0.444 &0.411 &0.611&0.635&0.655&\textbf{0.700} &&\\
    \hline
    $\alpha=3$ &MAP &0.515 &0.477 &0.674&0.689&0.713&\textbf{0.775}&&\\
    $C=10^3$ &MRR &0.201 &0.196 &0.212&0.234&0.224&\textbf{0.247} &50 &50,930\\
    $N$=500 &TOP-K &0.414 &0.401 &0.655&0.606&0.636&\textbf{0.711} &&\\
    \hline

    $\alpha=2$ &MAP &0.431 &0.518 &0.697&0.752&0.762&\textbf{0.810}&&\\
    $C=10^3$ &MRR &0.031 &0.036 &0.033 &0.045&0.039&\textbf{0.048} &500 &17,670\\
    $N$=5000 &TOP-K &0.401 &0.499 &0.663&0.702&0.711&\textbf{0.740} &&\\
    \hline
    $\alpha=2$ &MAP &0.455 &0.558 &0.722&0.777&0.797&\textbf{0.894}&&\\
    $C=10^4$ &MRR &0.038 &0.043 &0.045&0.050&0.047&\textbf{0.051} &500 &41,679\\
    $N$=5000 &TOP-K &0.423 &0.489 &0.678&0.711&0.737&\textbf{0.822} &&\\
    \hline
    $\alpha=3$ &MAP &0.471 &0.484 &0.708&0.769&0.787&\textbf{0.901}&&\\
    $C=10^3$ &MRR &0.039 &0.039 &0.046&0.048&0.046&\textbf{0.051} &500 &63,703\\
    $N$=5000 &TOP-K &0.442 &0.438 &0.682 &0.721&0.744&\textbf{0.841} &&\\
    \hline
    $\alpha=3$ &MAP &0.477 &0.498 &0.736&0.801&0.812&\textbf{0.907}&&\\
    $C=10^4$ &MRR &0.041 &0.042 &0.046&0.051&0.048&\textbf{0.052} &500 &94,612\\
    $N$=5000 &TOP-K &0.466 &0.467 &0.711&0.763&0.776&\textbf{0.837} &&\\
    \hline
  \bottomrule
\end{tabular}
\end{table*}

\subsection{Setup}

Our approach is based on random walk and dynamic Bernoulli embeddings, and there are several hyperparameters for the construction of dynamic networks, random walk, and embedding learning. We fix some hyperparameters (i.e., $D$=128, $L$=80, $r$=10, $cs$=4, $ns$=10) as suggested in \cite{CTDNE} and use the fixed event number strategy to construct dynamic networks for discrete-time embeddings methods (i.e., TNE, DynGEM, DynamicTriad and our method) in this section.

\subsection{Baseline Methods}

Our method is a discrete-time network embedding method, and we select the compared baseline methods from different categories. DeepWalk and node2vec are two representative static methods. For the evolving node detection task, we first apply them to learn each graph separately and use alignment methods~\cite{DWER} to align these embeddings into the same vector space. For dynamic network embedding methods, we select one continuous-time network embedding method (CTDNE) and three discrete-time network embedding methods (TNE, GynGEM, and DynamicTriad).

\begin{itemize}
\item \textit{DeepWalk}~\cite{deepwalk}. This static method is based on random walk and the skip-gram model. Three hyperparameters are set as default ($D=128$, $r=10$, $ns=10$) and the other two hyperparameters are selected from several values, $L\in\{40,60,80\}$, $cs\in\{6,8,10\}$.
\item \textit{node2vec}~\cite{node2vec}. node2vec captures the diversity of connectivity patterns in a network and preserves high-order proximities between nodes. node2vec introduces two new hyperparameters for grid search compared with DeepWalk and we set $p,q\in \{0.25,0.50,1,2\}$ as mentioned in \cite{node2vec}.
\item \textit{CTDNE}~\cite{CTDNE}. This is a continuous-time network embedding method based on DeepWalk to capture temporal information by random walk in the chronological order. We set $F_s$ as a linear distribution and $F_t$ as a unbiased distribution as suggested in \cite{CTDNE}.
\item \textit{TNE}~\cite{TNE}. This dynamic embedding method is based on matrix factorization to learn discrete-time network embeddings. We choose parameter $\lambda\in\{0.001,0.01,0.1,1,10\}$.
\item \textit{DynGEM}~\cite{DynGEM}. This model is based on deep auto-encoders to incrementally generates the embedding of the current graph with an initialization from the previous graph. We fix the parameters as suggested in~\cite{DynGEM}.
\item \textit{DynamicTriad}~\cite{triad}. This model utilizes \textit{triad} to capture the dynamic changes in networks. To report the best performance of this method, we set hyperparameters $\beta_1\in\{0.1,1,10\}$ and $\beta_2\in\{0.1,1,10\}$ alternatively.
\end{itemize}

We repeat all experiments 10 times and report the average performance of each method.

\subsection{Link Prediction}
\label{sec:link}

Link prediction is a common application to evaluate the performance of network embeddings. To generate the training data and testing data, we sort all temporal edges by the time-ascending order as suggested in \cite{CTDNE}. Then we use the first 75\% as the training data and the remaining 25\% as the testing data (one graph). Static methods (DeepWalk and node2vec) use the whole training data as one graph to learn one network embedding, while the training data is further partitioned into a sequence of graphs for TNE, DynGEM, DynamicTriad, and our method. The number of timesteps are listed in Table~\ref{tab:datasets}.

%with parameters $\varepsilon\in\{2000,4000,8000,16000\}$ and $\gamma\in{0,0.25,0.5,0.75}$

We compute the similarity between two nodes by the L2 distance in the current timestep to predict whether the two nodes exist one edge in the next timestep. To evaluate the performance by AUC, we use a \textit{logistic regression model} as the classifier with 5-fold cross-validation. Table 2 demonstrates that our method outperforms the other methods in most cases.

\subsection{Evolving Node Detection}
The structure of a real-world network changes over time. However, the network structure generally does not change sharply between adjacent timesteps and most nodes are stable in many timesteps. Dynamic network embeddings can be used to detect evolving nodes when their neighbors have changed significantly. Thus, we use the evolving node detection task to evaluate the proximity preservation and temporal continuity of network embedding methods in capturing evolution patterns.

For each timestep $t$, we calculate the distance between two embedded vectors $y_g^{(t)}$ and $y_g^{(t+1)}$ in adjacent timesteps for each node $g$, and sort these distances by the descending order. The first 10\% nodes with a large distance are called \textbf{active nodes} in the timespan $[S_t,S_{t+1}]$, and this timestep is an active timestep for these active nodes. We further sort active nodes in all timesteps by the number of active timesteps, and choose the top 10\% nodes with a large number of active timesteps as \textbf{evolving nodes}.

%A node $g$ has two embedding vectors $y_g^{(t)}$ and $y_g^{(t+1)}$ in adjacent timesteps, and we can compute the distance between $y_g^{(t)}$ and $y_g^{(t+1)}$, then sort all distances by the descending order and set first 10\% as a threshold. Therefore, nodes with a higher distance than the threshold are \textbf{active} in the timespan $[S_t,S_{t+1}]$. We sort all nodes by the times of being active in a descending order and pick the top 10\% as \textbf{evolving nodes}.

Table~\ref{tab:detection} shows the performance of six methods by three metrics: MAP, MRR and TOP-K. CTDNE generates one final embedding for a continuous-time network, and cannot be used for this task. Our method achieves overall the highest gains against other state-of-the-art methods.

\subsection{Parameter Analysis}

Many hyperparameters of our method have been evaluated by previous work~\cite{deepwalk,DBE}, and we select two hyperparameters (i.e., $\varepsilon$ and $\gamma$) from dynamic network construction and the other two hyperparameters (i.e., $r$ and $cs$) from network embedding learning for parameter analysis by evaluating the performance of our method in the link prediction task.

\textbf{Dynamic network construction}. To construct dynamic networks, we need to specify the event window size $\varepsilon$ and overlap ratio $\gamma$. Initially, we fix the other parameters (i.e., $D$=128, $L$=80, $r$=10, $cs$=4, $ns$=10), and then we evaluate the performance with different values of hyperparameters $\varepsilon$ and $\gamma$ by the link prediction task in two datasets, IA-ENRON and FB-FORUM. Table~\ref{tab:ana1} demonstrates the performance of our method in the link prediction task is stable with different values of $\varepsilon$ and $\gamma$.

\textbf{Network embedding learning}. We select two hyperparameters, $r$ (i.e., the number of walks of each node) from random walk and $cs$ (i.e., the context size) from Bernoulli embedding learning. We fix the other hyperparameters (i.e., $D$=128, $L$=80, $r$=10, $\varepsilon$=8000, $\gamma$=0.5) and change the values of $r$ and $cs$.

Table 5 shows the performance of our method is stable when $r=10$. However, if we fix $cs=4$, the best performance with $r=10$ achieves an average gain of 6.4\% across the other values of $cs$ for FB-FORUM. Thus, the hyperparameter $cs$ is a little sensitive for the link prediction task and we can tune the value of $cs$ to achieve better performance.

\begin{table}
\footnotesize
  \caption{AUC scores of link prediction for hyperparameter analysis of $\varepsilon$ and $\gamma$}
  \label{tab:ana1}
  \begin{tabular}{rcccc}
    \toprule
    $\varepsilon$&2000&4000&8000&16000\\
    \midrule
    IA-ENRON ($\omega$=0.5) &0.900 &0.919 &\textbf{0.930} &0.924\\
    FB-FORUM ($\omega$=0.5) &0.908 &0.910 &0.898 &\textbf{0.915}\\
    \bottomrule
    \\
    \toprule
    $\gamma$&0&0.25&0.50&0.75\\
    \midrule
    IA-ENRON ($\varepsilon$=8000) &0.933 &0.930 &\textbf{0.935} &0.920\\
    FB-FORUM ($\varepsilon$=8000) &0.901 &0.910 &\textbf{0.920} &0.895\\
    \bottomrule
\end{tabular}
\end{table}

\begin{table}
\footnotesize
  \caption{AUC scores of link prediction for hyperparameter analysis of $cs$ and $r$.}
  \label{tab:ana2}
  \begin{tabular}{rcccc}
    \toprule
    $cs$&2&4&8&16\\
    \midrule
    IA-ENRON ($r$=10) &0.920 &\textbf{0.934} &0.926 &0.905\\
    FB-FORUM ($r$=10) &0.920 &\textbf{0.921} &0.892 &0.884\\
    \bottomrule
    \\
    \toprule
    $r$&1&5&10&15\\
    \midrule
    IA-ENRON ($cs$=4) &0.925 &0.926 &\textbf{0.935} &0.934\\
    FB-FORUM ($cs$=4) &0.857 &0.870 &\textbf{0.920} &0.866\\
    \bottomrule
\end{tabular}
\end{table}

\section{APPLICATIONS}

To analyze evolution patterns of real-world dynamic networks, we visualize the trajectories of evolving nodes in 2D space by t-SNE~\cite{t-SNE}. A path with the color from light to dark (such as the light blue to the dark blue) shows the trajectory of an evolving node in a timespan and we draw the node every three timesteps. Note that the two dynamic networks in this section are regarded as undirected and we construct dynamic networks by the fixed time interval for evolution analysis.

\subsection{Primary school dynamic network}

\begin{figure}
  \subfigure[2D projection of network embeddings of students and teachers.]{
    \label{fig:4:a} %% label for first subfigure
    \includegraphics[width=8cm]{show_primary.pdf}
  }
  \subfigure[Number of evolving nodes in each timestep.]{
    \label{fig:4:b} %% label for second subfigure
    \includegraphics[width=8cm]{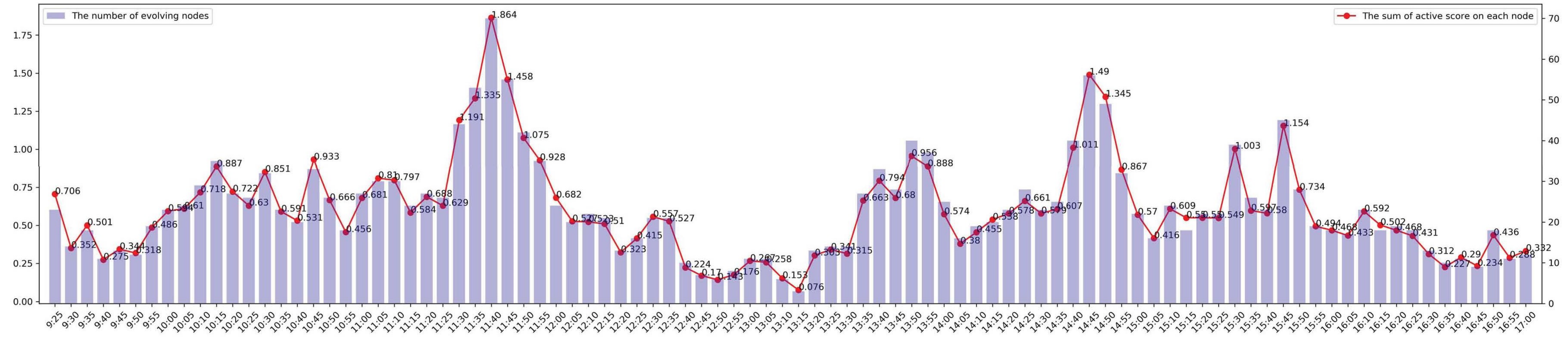}
  }
  \subfigure[Evolving nodes grouped by the number of active timesteps.]{
    \label{fig:4:c} %% label for second subfigure
    \includegraphics[width=8cm]{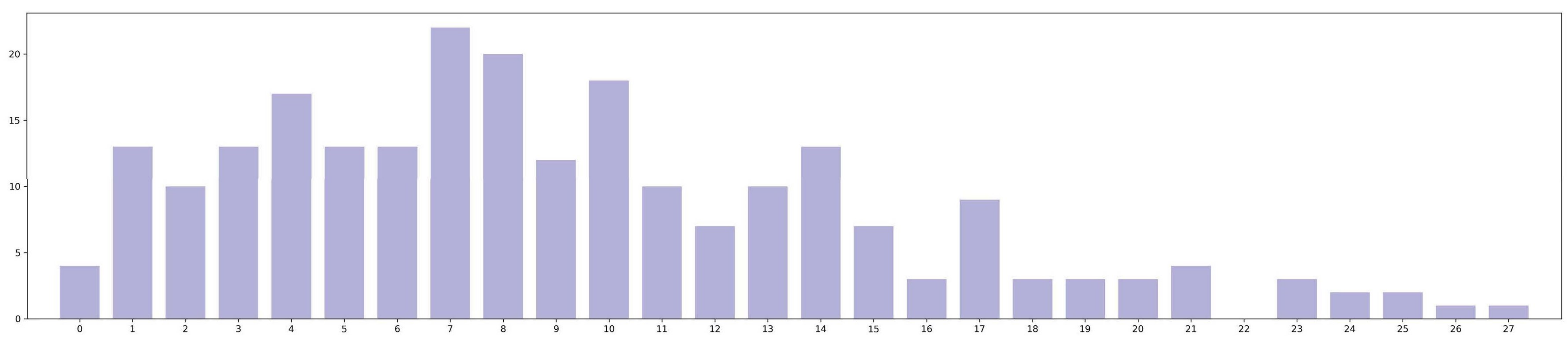}
  }
  \caption{Evolution analysis of the primary school dynamic network.}
  %\label{fig:subfig} %% label for entire figure
\end{figure}
\begin{figure}
  \centering
  \subfigure[]{
    \label{fig:3:a} %% label for first subfigure
    \includegraphics[width=5cm]{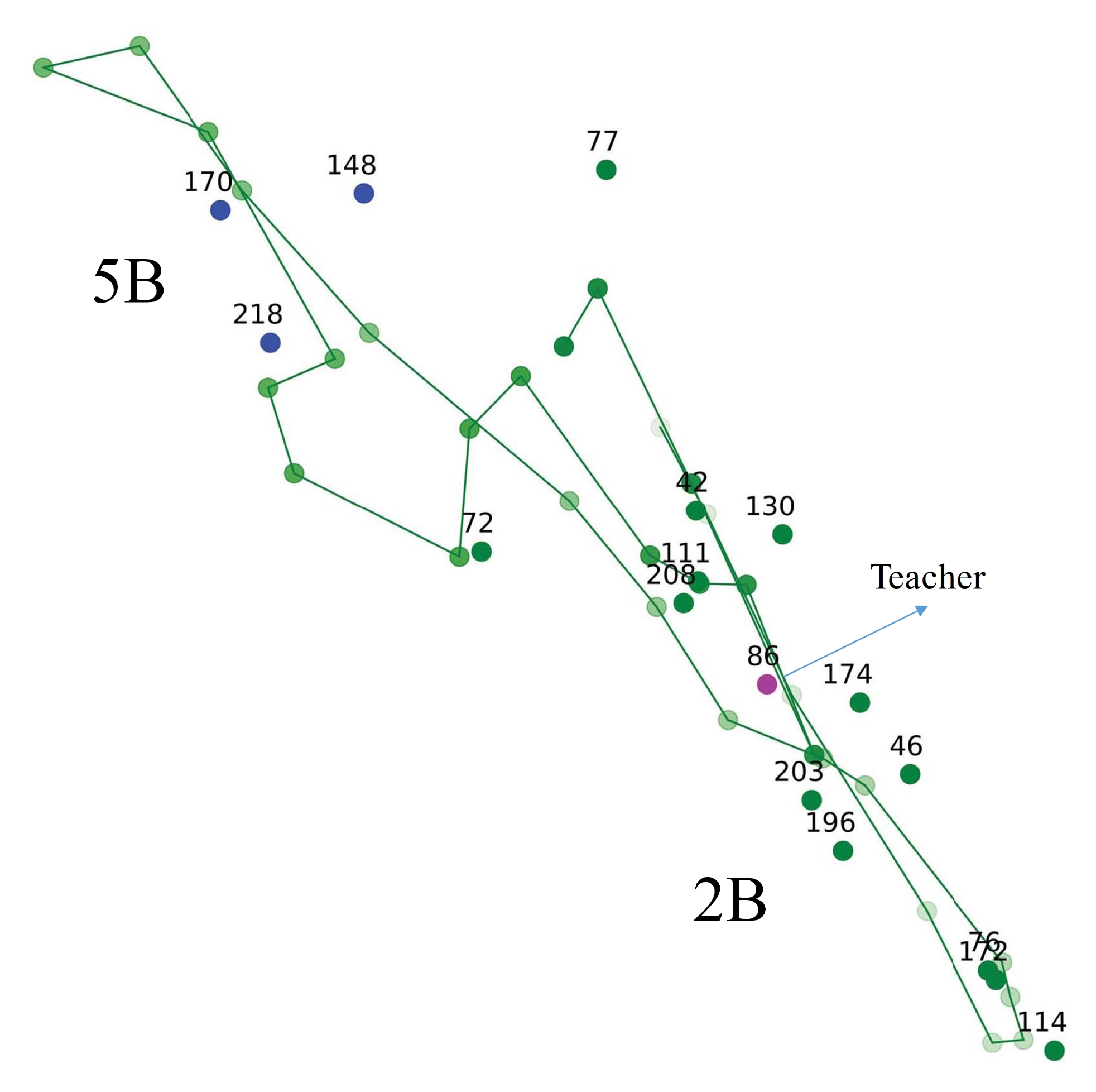}
  }
  \subfigure[]{
    \label{fig:3:b} %% label for second subfigure
    \includegraphics[width=5cm]{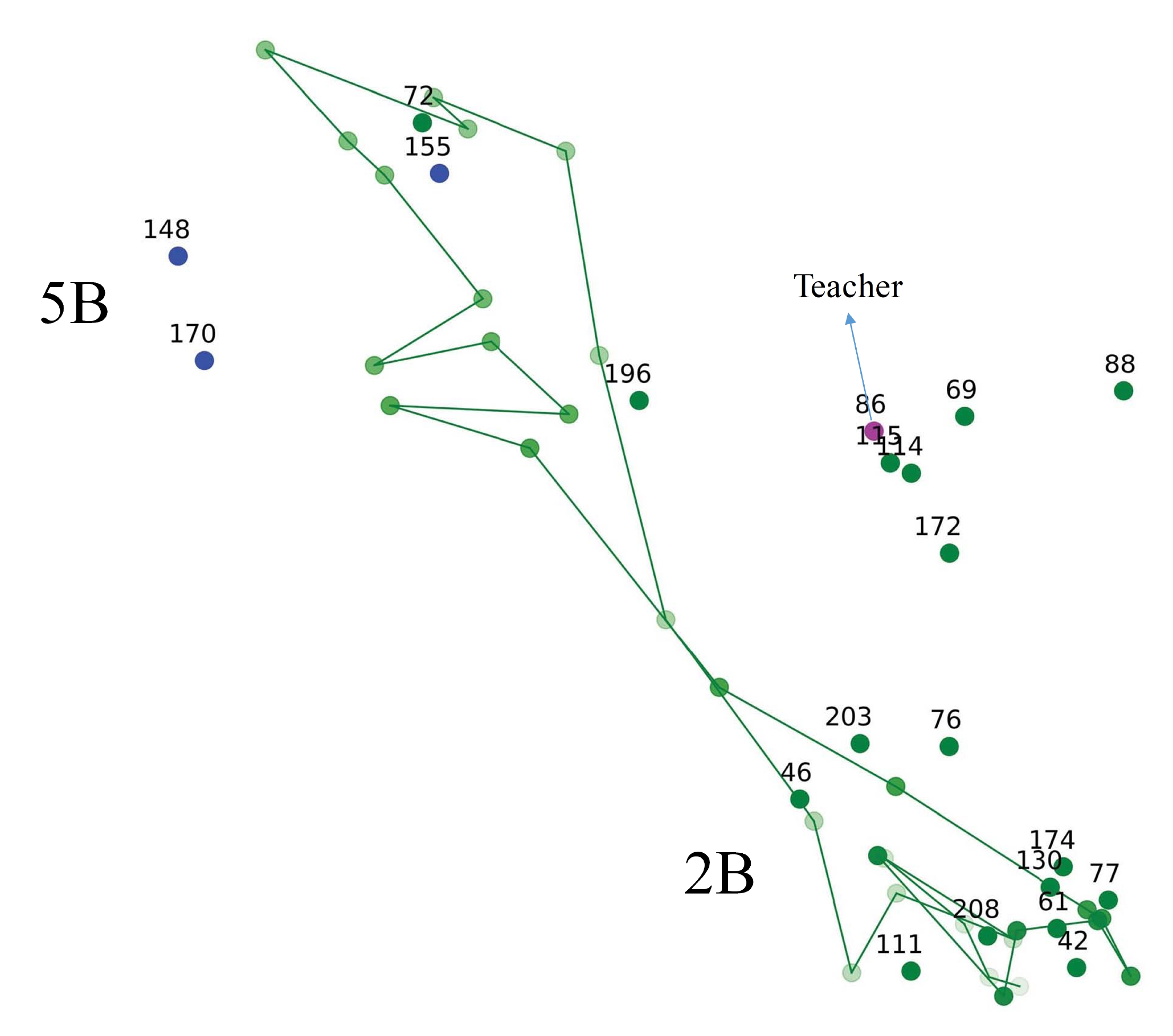}
  }
  \caption{The trajectories of Student 201 (a) and 74 (b).}
\end{figure}

The primary school data is collected from face-to-face contacts between students and teachers in a school locating in France during two school days in October 2009~\cite{HMOF}. This data contains 232 students from 10 classes, composed of 5 grades with each of the grades divided into two classes. There are 10 assigned teachers for 10 classes. We choose the first day to analyze its evolution patterns due to the similarity between two days. We create a dynamic network with 92 timesteps by a fixed time interval $\Delta t=6$ minutes and a window size $\omega=60$ minutes with an overlap ratio $\gamma=0.9$.

Figure~\ref{fig:4:a} shows the projection of students and teachers at 8:45 am (the first graph). We encode each class and teachers with different colors, total eleven colors. Figure~\ref{fig:4:b} depicts the number of evolving nodes defined in Section~\ref{sec:link} from 8:45 am to 17:05 pm. The evolution pattern is in line with the school schedule reported in~\cite{HMOF}. Two peak points at 11:40 am and 14:45 pm are two breaks and the low part shows the lunchtime from 12:30 pm to 14:00 pm. Furthermore, few students are extremely active as shown in Figure~\ref{fig:4:c}.

To analyze the trajectory of active students, we select Student 201 and Student 74 with higher active scores from 8:45 am to 17:05 pm in Figure 4. Student 201 and Student 74 move from Class 2B to Class 5B and then back to Class 2B. Student 201 contacts with the students of Class 5B (e.g., Students 148, 170 and 218) during the lunch break (i.e., from 11:30 am to 13:30 pm), thus we infer they may be friends. We may also conclude that Class 2B and Class 5B share the lunch break, which can be verified from the school schedule. Moreover, we notice that Student 201 contacts with Student 72 after the lunchtime, while Student 74 contacts with Student 72 before the lunchtime. This shows that Student 201 and Student 74 do not have a direct relationship. However, they may be both the friend of Student 72. Based on the triad theory, they will be friends via Student 72.

\begin{figure}
  \subfigure[2D projection of network embeddings colored according to its cluster membership.]{
    \label{fig:6:a} %% label for first subfigure
    \includegraphics[width=8cm]{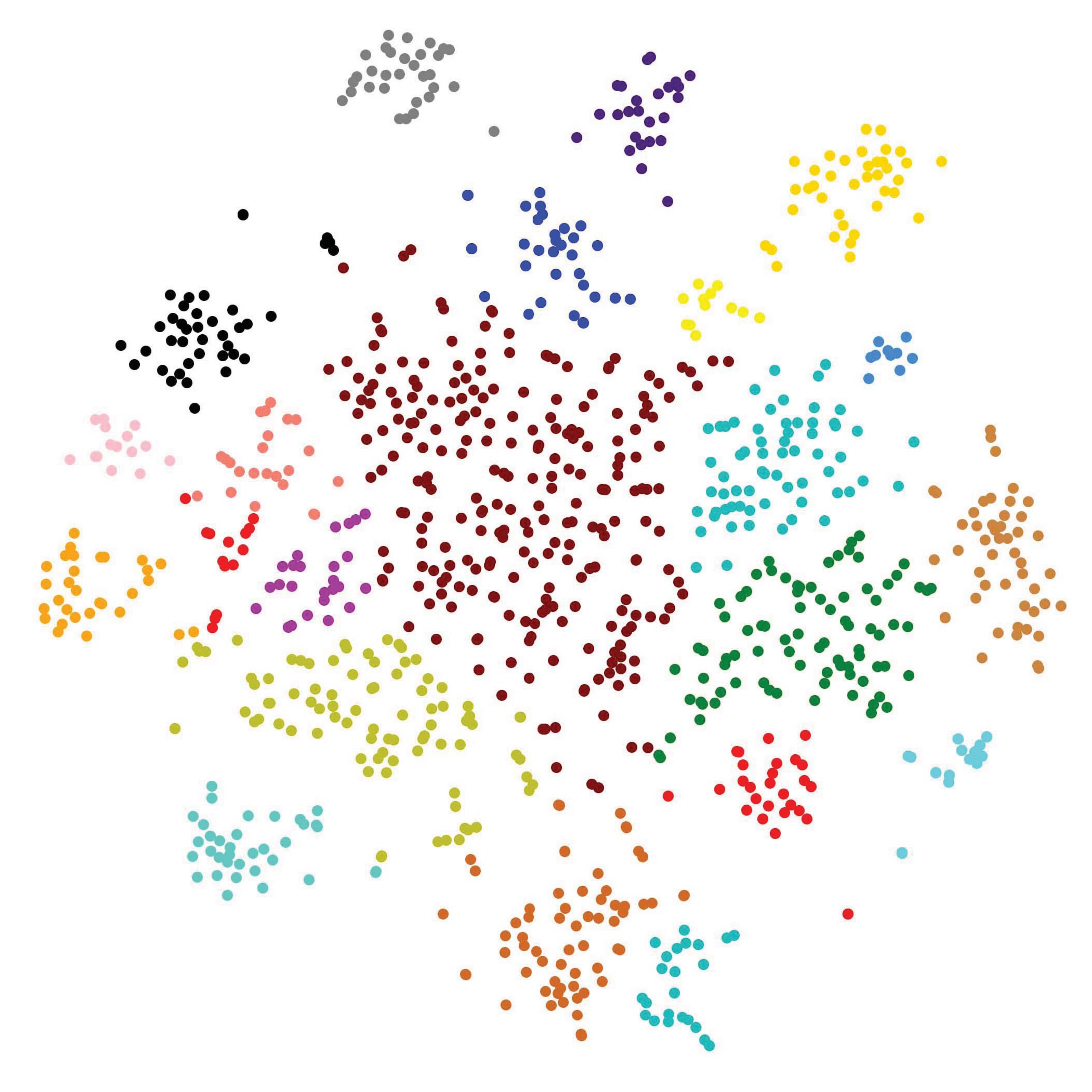}
  }
  \subfigure[Number of evolving nodes in each timestep.]{
    \label{fig:6:b} %% label for second subfigure
    \includegraphics[width=8cm]{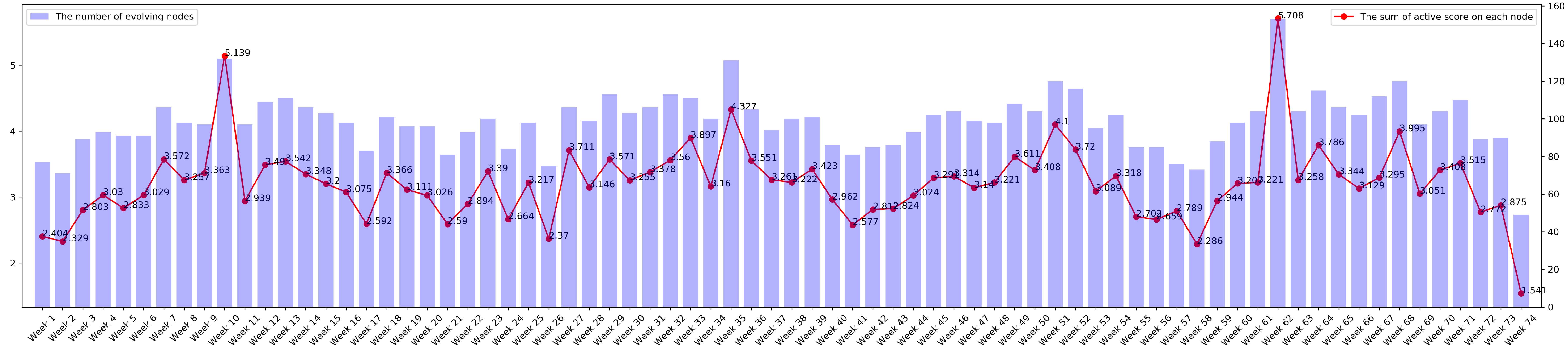}
  }
  \subfigure[Evolving nodes grouped by the number of active timesteps.]{
    \label{fig:6:c} %% label for second subfigure
    \includegraphics[width=8cm]{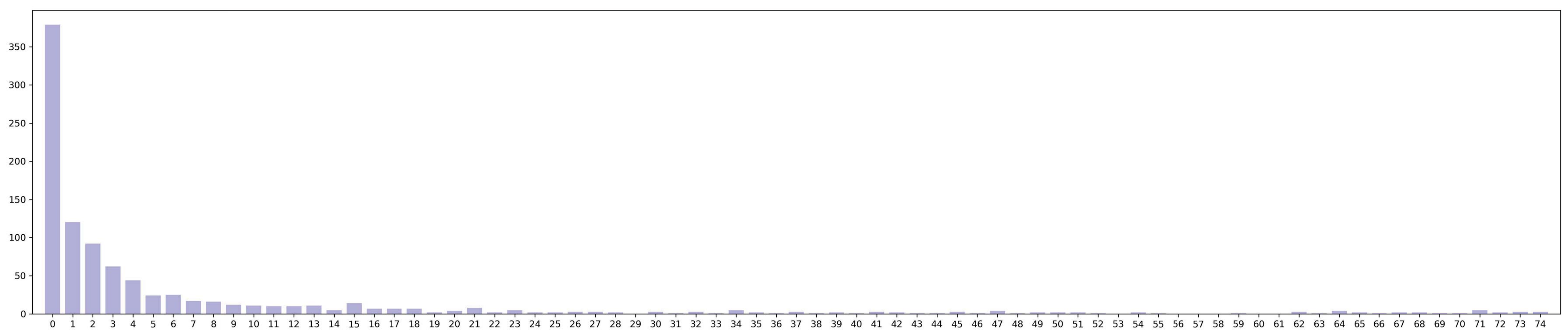}
  }
  \caption{Evolution analysis of the email communication dynamic network.}
  %\label{fig:subfig} %% label for entire figure
\end{figure}

\begin{figure}
  \centering
  \subfigure[]{
    \label{fig:5:a} %% label for first subfigure
    \includegraphics[width=5cm]{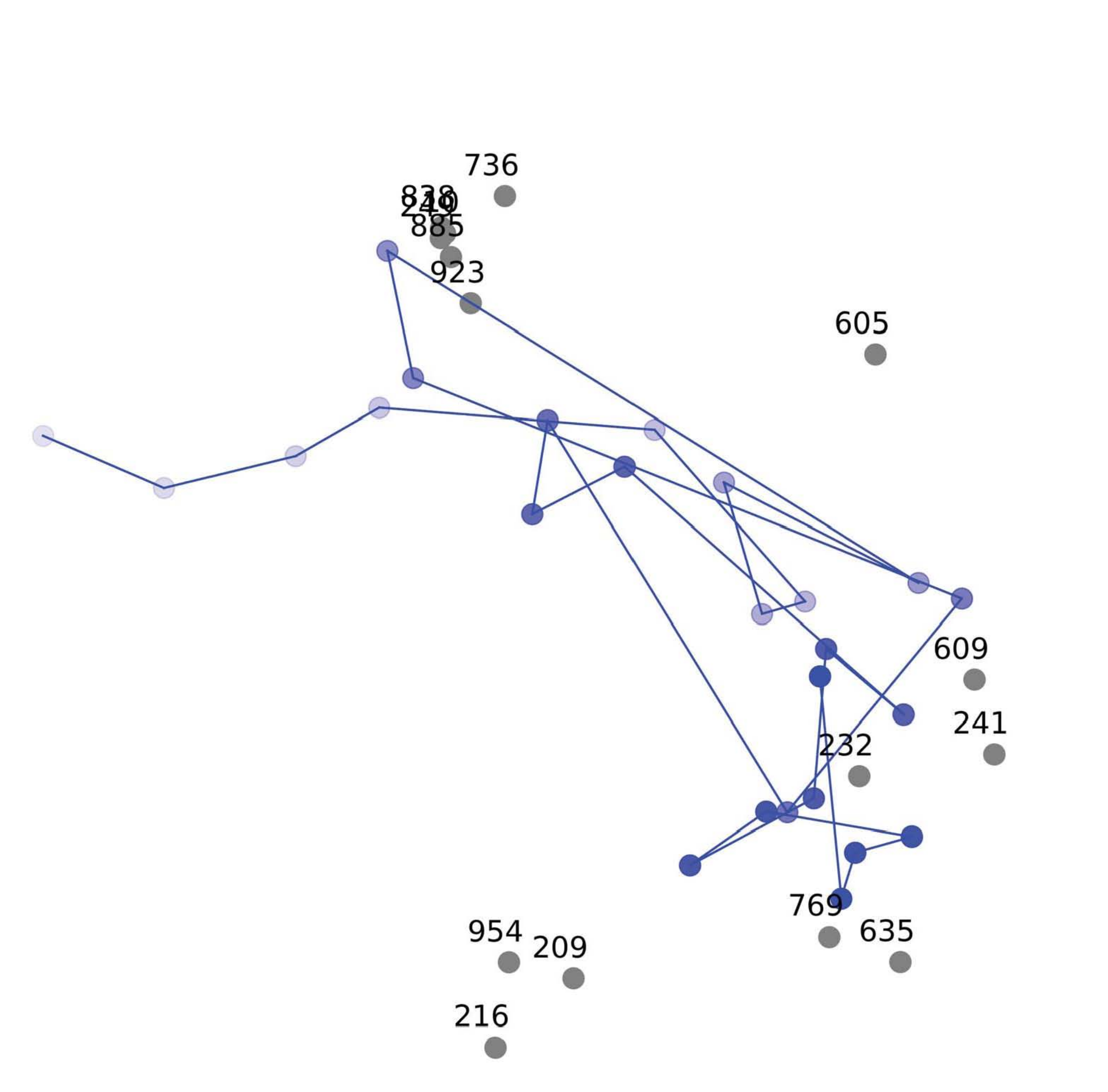}
  }
  \subfigure[]{
    \label{fig:5:b} %% label for second subfigure
    \includegraphics[width=5cm]{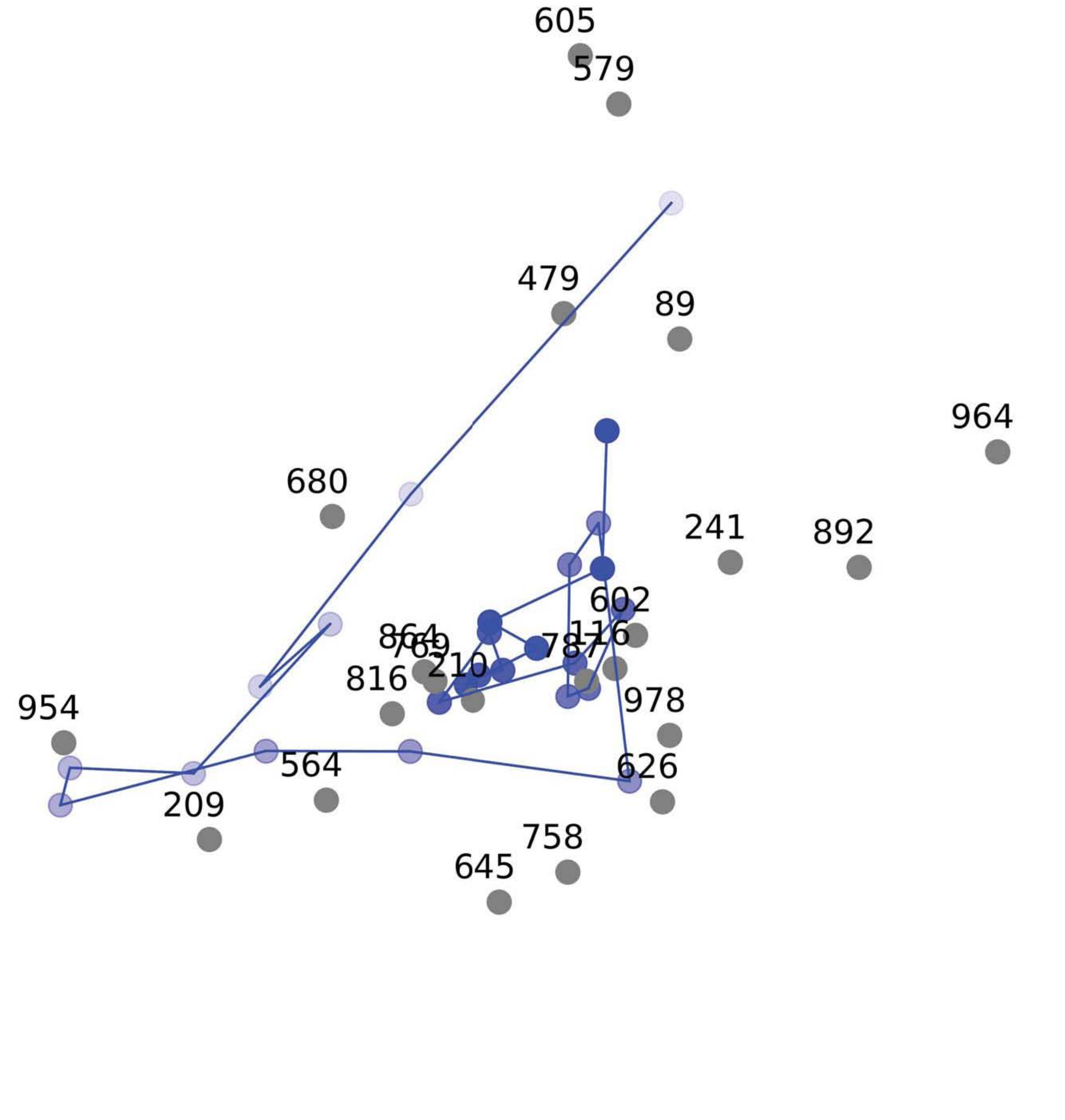}
  }
  \caption{The trajectories of Researcher 90 (a) and 328 (b).}
  %\label{fig:subfig} %% label for entire figure
\end{figure}

\subsection{Email communication dynamic network}

The email communication network data is collected from a large European research institution composed of 42 departments~\cite{MITN}. The network data contains 986 nodes and 332,334 temporal edges across 803 days. We create a dynamic network with 74 timesteps by a fixed time interval $\Delta t=7$ days and a window size $\omega=14$ days with an overlap ratio $\gamma=0.5$. The data provider hides all labels of departments for the consideration of personal privacy. To show the proximity of this network, we use DBSCAN~\cite{ADAF} to cluster nodes based on the L2 distance, and generate 42 clusters. Figure~\ref{fig:6:a} shows the clustering result at week 1 via t-SNE. Clusters are colored in different colors. As can be seen from Figure ~\ref{fig:6:a}, this network has a clear structure, and researchers may have more emails with each other in the same department.

%with parameters $eps=3.5$ and $samples=10$ to cluster nodes

The number of evolving nodes at each timestep is provided in Figure~\ref{fig:6:b}, and the number in each week is relatively stable except few weeks, such as week 10. Most nodes are stable (inactive), as most nodes have zero active weeks (timesteps) in Figure~\ref{fig:6:c}. We focus on two evolution trajectories of Researcher 90 and Researcher 328 for detailed analysis in Figure 6. Due to the lack of labels, neighbor nodes in the embedded space are colored in gray.

The neighbor nodes of Researcher 90 change over time and we notice Researcher 90 evolves between two groups. In the first few weeks, Researcher 90 has few neighbors, and the reason may be that Researcher 90 joined the institution as a new researcher. After week 51, Researcher 90 leaves his department to another.

Researcher 328 has few neighbors in the first few months. Researcher 90 and Researcher 328 have a similar evolution pattern at the beginning. After that, Researcher 328 moves from the outside of the department to the core position, while the contacts between Researcher 328 and others increases significantly. From the evolution trajectory of Researcher 328, he/she may become a leader or secretary in this department.

\section{CONCLUSION}

In this paper, we have proposed a novel approach to capturing the changes of dynamic networks with proximity and temporal properties preserved. To evaluate our method, we compare our method with several state-of-the-art methods including static methods and dynamic methods with two tasks, link prediction and evolving node detection. The experiments demonstrate that our method achieves substantial gains and perform effectively in proximity and evolution analysis of dynamic networks. For future work, it is desirable to update nodes which never appear in the network incrementally instead of retraining. Most existing network embedding methods only focus on one noticeable facet of the network, while the network includes diverse facets in the real world. Thus, we would like to design a method to incorporate multiple-facet properties into network embeddings.

\bibliographystyle{ACM-Reference-Format}
\bibliography{sample-bibliography}

\end{document}